\documentclass[epj, final]{svjour}

\usepackage{amssymb}
\usepackage{epsfig}

\newcommand{\mr}[1]{{\mathrm{#1}}}
\newcommand{\Tc}{\ensuremath{T_{\mathrm{c}}}}

\newcommand{\bfig}{\begin{figure}}
\newcommand{\efig}{\end{figure}}

\epsfclipon

\newcommand{\rsfig}[1]{\begin{center} 
                       \epsfig{file=#1, width=0.425\textwidth}
                       \end{center}
                       }

\begin{document}

\title{Glassy Dynamics of Simulated Polymer Melts:\\
Coherent Scattering and Van Hove Correlation Functions}
\subtitle{Part II: Dynamics in the $\alpha$-Relaxation Regime}

\titlerunning{$\alpha$-Dynamics of Glassy Polymer Melts}

\author{M.\ Aichele\inst{1}\fnmsep \thanks{Corresponding author.  E-mail: \textsf{Martin.Aichele@uni-mainz.de}} \and J.\ Baschnagel\inst{2}
} 

%
%

\institute{Institut f{\"u}r Physik Johannes Gutenberg-Universit{\"a}t Mainz, Staudinger Weg 7, 55099 Mainz, Germany \and Institut Charles Sadron, 6 rue Boussingault, 67083 Strasbourg, France}
\date{Received: date / Revised version: date}
%
\abstract{
Whereas the first part of this paper dealt with the relaxation in the $\beta$-regime, this part investigates the final relaxation ($\alpha$-relaxation) of a simulated polymer melt consisting of short non-entangled chains in the supercooled state above the critical temperature $\Tc$ of ideal mode-coupling theory (MCT).  The temperature range covers the onset of a two-step relaxation behaviour down to a temperature merely 2\% above $\Tc$.  We monitor the incoherent intermediate scattering function as well as the coherent intermediate scattering function of both  a single chain and the melt over a wide range of wave numbers $q$.  Upon approaching $\Tc$ the coherent $\alpha$-relaxation time of the melt increases strongly close to the maximum $q_\mr{max}$ of the collective static structure factor $S_q$ and roughly follows the shape of $S_q$ for $q \gtrsim q_\mr{max}$.  For smaller $q$-values corresponding to the radius of gyration the relaxation time exhibits another maximum.  The temperature dependence of the relaxation times is well described by a power law with a $q$-dependent exponent in an intermediate temperature range.  Deviations are found very close to and far above $\Tc$, the onset of which depends on $q$.  The time-temperature superposition principle of MCT is clearly borne out in the whole range of reciprocal vectors.  An analysis of the $\alpha$-decay by the Kohlrausch-Williams-Watts (KWW) function reveals that the collective KWW-stretching exponent and KWW-relaxation time show a modulation with $S_q$.  Futhermore, both incoherent and coherent KWW-times approach the large-$q$ prediction of MCT already for $q > q_\mr{max}$.  At small $q$, a $q^{-3}$-power law is found for the coherent chain KWW-times similar to that of recent experiments.
} 

\maketitle



\section{Introduction}
\label{sec:intro}
The preceding paper \cite{betaDynamics} reported results from a molecular-dynamics simulation for a bead-spring model of a supercooled polymer melt.  The aim of this work was to extend our previous analysis \cite{BennemannPaulBinder1998,BennemannPaulBaschnagel1999,BennemannBaschnagelPaul1999_incoherent,BennemannPaulBaschnagel1999_Rouse,natureBDBG1999,BBPB_2000} to coherent scattering of the chains and the melt.  Much of this analysis was guided by the mode-coupling approach to the structural glass transition \cite{Goetze1999_review,goetzemctessentials,GoetzeSjoegren1995_TTSP,Goetze_LesHouches}.  

Mode-coupling theory (MCT) derives an equation of motion for density fluctuations (incoherent and coherent scattering functions), in which a non-linear coupling between different wave vectors (``modes'') gives rise to a critical temperature $\Tc$.  This temperature is situated in the region of the supercooled liquid above the glass transition temperature $T_\mr{g}$.  It may be interpreted as a crossover point from the high-temperature region of structural relaxation (dominated by the ``cage effect'') to the low-tem\-pe\-ra\-ture region where the dynamics becomes more and more dominated by activated processes over (free) energy barriers during supercooling (see \cite{goetzemctessentials,GoetzeSjoegren1995_TTSP} for a detailed discussion of this physical picture).  

An important property of $\Tc$ is that it is an equilibrium quantity which can be calculated if accurate data for the static structure factor are available.  Such a quantitative approach was performed for hard-sphere-like colloidal particles \cite{GoetzeSjoegren1991_colloids,vanMegen1995}, soft spheres \cite{BarratLatz1990} or binary Lennard-Jones (LJ) mixtures \cite{NaurothKob1997}.  These approaches reveal that MCT provides a reasonable framework for quantitative predictions of specific systems, but it also tends to overestimate the freezing ability of a glass former (see \cite{NaurothKob1997} for a comparative discussion). 

This tendency is exemplified by the so-called idealized MCT which deals with the cage effect only \cite{goetzemctessentials,GoetzeSjoegren1995_TTSP,Goetze_LesHouches}.  The idealized theory predicts the intermittence of complete structural relaxation if $T \leq \Tc$.  However, since $\Tc > T_\mr{g}$, the  arrest of structural relaxation is in general not observed in experiments \cite{Goetze1999_review,Vigo1997,Pisa1998} or computer simulations \cite{Goetze1999_review,kobreview1999}.  Additional relaxation channels, which are ignored by the idealized MCT, must become important close to and particularly below $\Tc$.  The extended MCT tries to include these channels approximately \cite{GoetzeSjoegren1995_TTSP,FuchsGoetzeHildebrand1992_extMCT,GoetzeSjoegren1987_extMCT}.  These additional relaxation processes, called ``activated hopping processes'', replace the complete freezing of the idealized theory at $\Tc$ by the aforementioned crossover to low-temperature activated dynamics.  The status of this extension is, however, unclear at present \cite{Goetze1999_review,goetzevoigtmann2000}. On the one hand, evidence from simulations is scarce \cite{BaschnagelFuchs1995}, and on the other hand, evidence from experiments \cite{Cummins_extMCT1,Cummins_extMCT2,Cummins_extMCT3} is challenged due to experimental problems \cite{flaw_in_Fabry_Perot}.  

Nonetheless, a recent application of a schematic MCT model, including hopping processes, to light-scattering, dielectric relaxation and neutron scattering data \cite{goetzevoigtmann2000} yields a coherent description of the different spectra.  This study also exemplifies the theoretical prediction \cite{FuchsGoetzeHildebrand1992_extMCT} that there is a temperature interval above $\Tc$, where the idealized theory represents a viable approach.  The same conclusion was also drawn from various other experiments \cite{Goetze1999_review,vanMegen1995,vanMegenMortensen1998,Bartsch1998,ToelleWuttkeSchober1998,ToelleSchoberWuttke1997,LunkenheimerReview2000,HinzeBrace2000} and simulations \cite{kobreview1999,KaemmererKobSchilling1998,KaemmererKobSchilling1998_orient,GleimKob2000,HorbachKob1999,Mossa2000,TheisSciortino2000,sciortino2000,ZonLeeuw1998}.  This has led to extension of the idealized theory to include corrections to the asymptotic behavior close to $\Tc$ \cite{FranoschFuchsGoetze1997,FuchsGoetzeMayr1998} or to orientational degrees of freedom \cite{SchillingScheidsteger1997,FabbianLatz2000,FranoschGoetze_orient1997}, but also to further tests of the theory.

In this spirit, we want to complement our previous simulation studies by an analysis of the $\alpha$-relaxation of the coherent scattering functions of a supercooled polymer melt.  This analysis is based on a detailed investigation of the $\beta$-relaxation described in the preceding paper \cite{betaDynamics}.  The present paper is organized as follows:  Section~\ref{sec:model} introduces the model and the analysed quantities, whereas section~\ref{sec:theory} reviews the theoretical background for the subsequent analysis which is described in Section~\ref{sec:results}.  The final section contains our conclusions.

\section{Model and Analysed Quantities}
\label{sec:model}
The model we simulated by means of molecular-dynamics simulation is described in part I of this paper \cite{betaDynamics} (hereafter referred to as part I) and more extensively in \cite{BennemannPaulBinder1998}.  Here, we only give a brief summary.

We simulated a bead-spring model of linear polymers with $N=10$ monomers per chain.  The monomer-monomer interaction was modelled by a truncated and shifted Len\-nard-Jones (LJ) potential.  So, all quantities are measured in LJ-units in the following.  In addition, bonds along the polymer backbone were introduced by a FENE-potential.  The parameters of the potentials were adjusted such that the minima of the LJ-potential and of the bond potential (= combined LJ- and FENE-potentials), $r_\mr{min}$ and $r_\mr{bond}$, respectively, were geometrically incompatible.  This means that the preferred nearest-neighbour distance of non-bonded monomers $r_\mr{min}$ and the preferred bond length $r_\mr{bond}$ do not allow arrangement of the beads in a regular fcc (or bcc) lattice structure.  Thus, cristallisation is prevented by this incompatibility and by the flexibility of the polymer backbone at all temperatures (the (square) end-to-end distance $R_\mr{e}^2 = 12.3 \pm 0.1$ and the radius of gyration $R^2_\mr{g} = 2.09 \pm 0.01$ stay almost constant \cite{BennemannPaulBinder1998}).

Simulations were done at constant pressure $p=1$ and at temperatures $T=0.46$, 0.47, 0.48, 0.50, 0.52, 0.55, 0.6, 0.65, 0.7, covering the range from the onset of glassy behaviour to about the critical temperature of mode-coupling theory (MCT), $\Tc = 0.450 \pm 0.005$ (see \cite{BennemannPaulBinder1998} and section~\ref{subsec:reltimes}).  For all temperatures an amorphous static structure is observed, as discussed in part I.  

Whereas the first part of the paper investigated the behaviour of our system in the $\beta$-relaxation regime, this part discusses the $\alpha$-relaxation, i.e., the late-time structural relaxation of the polymer melt.  We computed the incoherent, the coherent chain and the coherent melt intermediate scattering functions to probe the dynamics.

We define the coherent (or collective) intermediate scattering function of the melt by
\begin{equation}
\label{eq:def_phiqt}
\phi_q(t) := \frac{S_q(t)}{S_q(0)} \;,
\end{equation}
where $S_q(t)$ is given by
\begin{equation}
\label{eq:defsqt}
S_q(t) := \frac{1}{M} \Bigl\langle \sum_{i=1}^M\sum_{j=1}^M \exp \left\{ \mr{i} \vec{q}\cdot\bigl[\vec{r}_i(t)-\vec{r}_j(0)\bigl]\right\} \Bigl\rangle \;.
\end{equation}
Monomer $i$ has coordinate $\vec{r}_i(t)$ at time $t$, $M$ is the total number of monomers in the melt, and $\langle \cdot \rangle$ denotes the canonical ensemble average.  $S_q(0) \equiv S_q$ is the melt's static structure factor.  For defining the coherent chain scattering function $\phi^{\mr{p}}_q(t)$ we have to use $N$ instead of $M$ as upper summation index in equation~(\ref{eq:defsqt}) and sum over all pairs of monomers in a polymer.  If we set $i=j$ in equation~(\ref{eq:defsqt}), we obtain the single particle incoherent scattering function $\phi^\mr{s}_q(t)$.  A superscript ``s'' for incoherent (self) and a superscript ``p'' for chain (polymer) quantities will be used in the following, while coherent quantities do not carry a superscript.  When referring to all scattering functions, we shall frequently use the notation $\phi^\mr{x}_q(t)$. The superscript ``x'' denotes the dependence on the correlators and is used analogously for other quantities.  Further details about the computation of the scattering function may be found in part I.  In our isotropic system all quantities depend on the modulus $q = |\vec{q}|$ of the reciprocal vector $\vec{q}$ only.

\section{Theoretical Background}
\label{sec:theory}
This section briefly  summarizes predictions of the idealized mode-coupling theory (MCT) for the $\alpha$-process (see e.g.\ \cite{Goetze1999_review,Goetze_LesHouches,kobreview1999}), which we will apply in the analysis of the simulation data.  The relevant temperature scale of the theory is defined by the separation parameter $\sigma$,
\begin{equation}
\sigma := C \,\frac{\Tc-T}{\Tc} \; .
\end{equation}
Here, $C$ is a  system dependent constant. The separation parameter determines the temperature dependence of the universal $\alpha$-time scale by  
\begin{equation}
\label{eq:tau_alpha}
\tilde{\tau} := \frac{t_0}{|\sigma|^{\gamma}}\;, \quad T \geq \Tc \;,
\end{equation}
where $t_0$ is the microscopic matching time of MCT. For our system, the critical temperature $\Tc$ and the exponent $\gamma$ of the $\alpha$-process are given by $\Tc=0.450\pm 0.005$ and $\gamma=2.09\pm 0.07$ \cite{BennemannBaschnagelPaul1999_incoherent}.

Mode-coupling theory predicts that the correlator $\phi_q^\mr{x}(t)$ decays in two steps in the vicinity of $\Tc$.  In the first step $\phi_q^\mr{x}(t)$ relaxes towards a plateau value, the non-ergodicity parameter $f_q^\mr{x c}$ ($0 < f_q^\mr{x c} < 1$), whereas the second step represents the first part of the final structural relaxation to 0 ($\alpha$-process). 

An important prediction for the $\alpha$-process is the time-temperature superposition principle (TTSP).  This means that correlators $\phi^\mr{x}_q(t)$, measured at different temperatures, fall onto a master curve in the limit $\sigma \rightarrow 0$ when time is rescaled by the $\alpha$-relaxation time $\tau^\mr{x}_q$, defined by the decay of $\phi^\mr{x}_q(t)$ to a certain value $d$,
\begin{equation}
\label{eq:def_tau_alpha}
\phi^\mr{x}_q(\tau^\mr{x}_q)=d \;.
\end{equation}
The TTSP implies that the long-time relaxation of $\phi^\mr{x}_q(t)$ may be expressed as
\begin{equation}
\label{eq:ttsp_phi}
\phi^\mr{x}_q(t,T) = \tilde{\phi}^\mr{x}_q\left[t/\tau^\mr{x}_q(T)\right] \;.
\end{equation}
Furthermore, the $\alpha$-relaxation time $\tau^\mr{x}_q$ is predicted to factorize into two parts: into a correlator (x) and wave vector ($q$) dependent part $C^\mr{x}(q,d)$ and into the universal $\alpha$-time $\tilde{\tau}$,
\begin{equation}
\label{eq:ttsp_tau}
\tau^\mr{x}_q(T) = C^\mr{x}(q,d) \tilde{\tau}(T) = C^\mr{x}(q,d) \frac{t_0}{|\sigma|^\gamma}\;.
\end{equation}
Thus, all relaxation times $\tau^\mr{x}_q$ in the $\alpha$-regime exhibit the same temperature dependence.  This property is in fact the reason why $\tau^\mr{x}_q$ may be defined by $\phi(\tau^\mr{x}_q) = d$, with ``$d$'' being any arbitrary number sufficiently smaller than the non-ergodicity parameter $f_q^\mr{xc}$ so that $\phi(\tau^\mr{x}_q)$ probes the $\alpha$-process.  In our analysis, we chose $d = 0.1$, which satisfies this condition well for all $q$-values under investigation.  A different choice of $d$ would only alter $C^\mr{x}(q,d)$, but not the temperature dependence.  

The short-time expansion of the correlators in the $\alpha$-regime is given by \cite{FranoschFuchsGoetze1997}
\begin{equation}
\label{eq:early_alpha}
\phi^\mr{x}_q(t) = f^\mr{xc}_q - h^\mr{x}_q B \left(t/\tilde{\tau}\right)^b \;,
\end{equation}
with the von Schweidler exponent $b=0.75 \pm 0.04$ and $B=0.476 \pm 0.060$ \cite{BennemannBaschnagelPaul1999_incoherent} for our system.  The critical amplitude $h^\mr{x}_q$ and $f^\mr{xc}_q$ are investigated in detail in part I.

In experiments and simulations one often finds that a Kohlrausch-Williams-Watts
(KWW) function, defined by  
\begin{equation}
\label{eq:KWW-function}
\Phi^\mr{x K}_q(t) = f^\mr{x K}_q \exp \! \left[-(t/\tau^\mr{x K}_q)^{\beta^\mr{x K}_q}\right] \;,
\end{equation}
gives a very good description of the correlator in the $\alpha$-regime.  However, especially in the crossover region from the $\beta$- to the $\alpha$-regime small systematic deviations are commonly reported \cite{Goetze_LesHouches,BoehmerNgaiAngellPlazek1993,FuchsGoetzeHofackerLatz1991,FuchsHofackerLatz1992}.  Mode-coupling theory rationalizes these deviations by the fact that the short-time expansion of equation~(\ref{eq:KWW-function}) does not coincide with equation~(\ref{eq:early_alpha}) because $b \neq \beta^\mr{x K}_q$ in general \cite{FuchsHofackerLatz1992}. However, in the limit of large $q$ it was proved \cite{Fuchs1994_kww} that all correlators exhibit the following KWW behaviour,
\begin{equation}
\label{eq:kww_limit}
 \lim_{q \rightarrow \infty} \phi^\mr{x}_q(t) = f^\mr{xc}_q \exp \! \left[-\Gamma^\mr{x}_q(t/\tilde{\tau})^b\right], \; \Gamma^\mr{x}_q \propto q \;.
\end{equation}
So, $f^\mr{x K}_q = f^\mr{xc}_q$ in this limit.  On the other hand, the short-time limit of equation~(\ref{eq:early_alpha}), which is valid for any $q$, suggests that the KWW-prefactor should always be equal (or at least close) to $f^\mr{xc}_q$.  Another consequence of Eq.~(\ref{eq:kww_limit}) is that the stretching exponent $\beta^\mr{x K}_q$ satisfies
\begin{equation}
\label{eq:limit_beta}
\lim_{q \rightarrow \infty}\beta^\mr{x K}_q = b \;,
\end{equation}
and that the $q$-dependence of the KWW-time-scales $\tau^\mr{x K}_q$ is given by
\begin{equation}
\label{eq:limit_tau_kww}
\lim_{q \rightarrow \infty} \tau^\mr{x K}_q  = \left(\Gamma^\mr{x}_q\right)^{-1/b} \tilde{\tau} \propto q^{-1/b} \tilde{\tau} \;.
\end{equation}
The predictions presented in this section are valid in leading order of $\sigma$, corrections are of order $|\sigma|$.  This is the reason why the $\alpha$-scaling should extend to higher temperatures than the $\beta$-scaling, corrections to which are of order $\sqrt{|\sigma|}$ \cite{Goetze1999_review,FranoschFuchsGoetze1997,FuchsGoetzeMayr1998}.  

On the other hand, deviations are in general expected at high temperatures, where $\sigma$ is large, and at temperatures close to $\Tc$, where relaxation processes, which are not taken into account in the idealized MCT, begin to dominate the dynamics \cite{FuchsGoetzeHildebrand1992_extMCT}.

\section{Results}
\label{sec:results}

\subsection{$\alpha$-Relaxation Times}
\label{subsec:reltimes}

\bfig
\rsfig
{figures/tau_q_and_S_q_vs_q_T046_T052_T07.eps}
\caption[]{
\label{fig:tau_q_and_S_q_vs_q_T046_T052_T07}
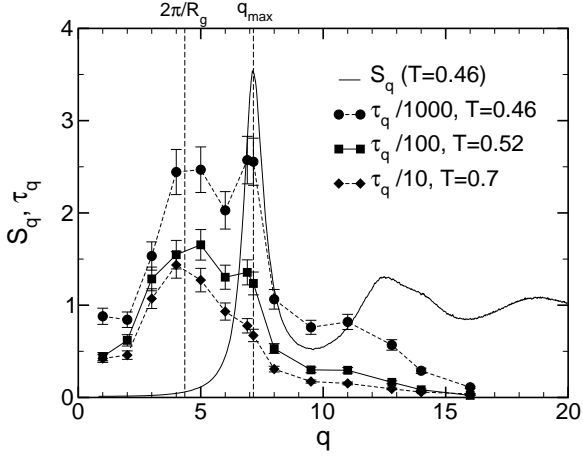
Static structure factor $S_q$ and collective $\alpha$-relaxation times $\tau_q(T)$ (scaled) for $T=0.46$, $T=0.52$, and $T=0.7$ versus $q$.  Vertical long dashed lines are drawn at $2\pi /R_\mr{g} = 4.35$ and at $q_\mr{max}\approx 7.15$, the maximum of $S_q$.  $S_q$ does not change much in this temperature interval (see part I). So, it is shown for $T=0.46$ only.
}
\efig

The $\alpha$-relaxation times $\tau^\mr{x}_q$ can be read off from the simulated scattering functions by determining the time when $\phi^\mr{x}_q(\tau^\mr{x}_q) = 0.1$ (see Eq.~(\ref{eq:def_tau_alpha})).  This means that the $\alpha$-decay times are obtained independently of any other quantities, such as $f^\mr{x K}_q$ or $\beta^\mr{x K}_q$.  Figure~\ref{fig:tau_q_and_S_q_vs_q_T046_T052_T07} shows the collective $\alpha$-relaxation time $\tau_q$ and the static structure factor $S_q$ as a function of $q$.  The comparison of both quantities reveals two conspicuous features: First, as temperature approaches $\Tc$, $\tau_q$ develops modulations which closely resemble the oscillations of $S_q$ if $q\gtrsim q_\mr{max}$.  This is particularly prominent at the maximum of the structure factor and also visible around the first minimum of $S_q$.  It means that the local environment around a particle determines the relaxation more and more as temperature approaches $\Tc$.  Such an adjustment of the $q$-dependence of $\tau_q$ to that of $S_q$ for $T \rightarrow \Tc$ is predicted by mode-coupling theory for hard spheres \cite{FuchsHofackerLatz1992} and also observed in other simulations \cite{KaemmererKobSchilling1998,KobAndersen_LJ_II_1995,SciortinoFabbianChen1997} and experiments on non-polymeric glass formers \cite{ToelleWuttke_EPJB1998,MezeiKnaakFarago1987}. In this respect, the considered polymer melt behaves similarly to simple liquids.  Moreover, $\tau_q$ exhibits an additional characteristic feature.  For all temperatures there is a broad peak around $q \approx 4.5$, which can be interpreted as corresponding to the radius of gyration $R_\mr{g}$ (reciprocal length $2\pi/R_\mr{g} = 4.35$).  A similar peak at a wave vector corresponding to the size of a molecule may also be observed in a simulation of a diatomic liquid \cite{KaemmererKobSchilling1998}. 

Although the abovementioned features are clearly visible in our simulation, we are not aware of similar findings in experiments on glass forming polymer melts.  We want to postpone a discussion of this point to Section~\ref{subsec:kww}, where we show results on KWW-fits.  The KWW-function is also used for the description of the $\alpha$-decay in experiments \cite{ArbeRichterColmenero1996,RichterMonkenbuschAllgeier1999} so that a qualitative comparison is better possible than for $\tau_q$.

Figure~\ref{fig:log_tau_alpha_chain+melt_vs_log_T-T_c} tests the prediction of equation~(\ref{eq:ttsp_tau}) for the $\alpha$-relaxation times of the coherent scattering functions.  It shows $\tau_q^\mr{p}$ and $\tau_q$ versus $T-\Tc$ in a double logarithmic plot, where $\Tc$ is kept fixed at $\Tc=0.450$.  In this representation, MCT predicts the simulation data to fall on straight lines with slope $\gamma=2.09$ for all relaxation times, irrespective of the $q$-values and the correlator considered.  This expectation is not fully borne out.  The data are linearized in an intermediate temperature interval only. Deviations appear for all $q$-values if $T-\Tc \lesssim 0.02$, and for large wave vectors if $T-\Tc \gtrsim 0.1$. These deviations are not unusual. They are also observed in other simulation studies \cite{KaemmererKobSchilling1998,KobAndersen_LJ_II_1995} and can be rationalized as follows:  Equation~(\ref{eq:ttsp_tau}) is only valid asymptotically close to the critical point so that deviations must occur as $T-\Tc$ grows.  The onset of these deviations depends on the correlator and on $q$.  Large $q$ probe small distances in real space where microscopic details should be felt more strongly.  So, it is possible that the temperature region of applicability is left earlier at large $q$.  On the other hand, very close to $\Tc$ the idealized MCT is no longer expected to be generally applicable because it neglects certain relaxation channels (called ``hopping processes'' \cite{FuchsGoetzeHildebrand1992_extMCT,GoetzeSjoegren1987_extMCT,GoetzeSjoegren1988_extMCT,Sjoegren1990_extMCT}), which are present in (almost) all real systems (some colloidal suspensions being an exception; see \cite{vanMegen1995,vanMegenMortensen1998} or \cite{Bartsch1998} for a recent review). These processes allow the correlators to decay  even below $\Tc$, where the ideal MCT would have predicted a cease of complete structural relaxation \cite{Goetze1999_review,Goetze_LesHouches}. 

\bfig
\rsfig{figures/log_tau_alpha_chain_vs_log_T-T_c.eps}
\rsfig{figures/log_tau_alpha_melt_vs_log_T-T_c.eps}
\caption[]{
\label{fig:log_tau_alpha_chain+melt_vs_log_T-T_c}
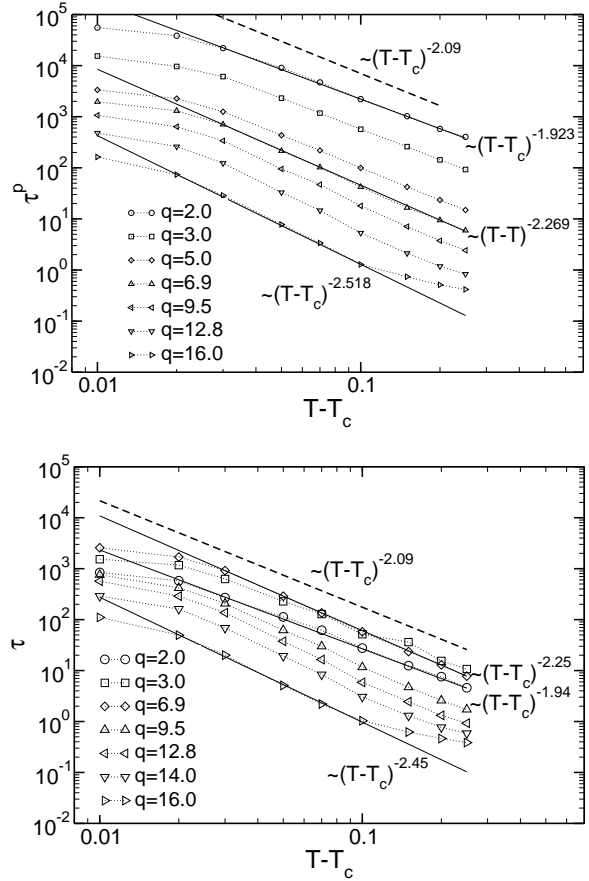
$\alpha$-relaxation times of the chain, $\tau^\mr{p}_q$ (top), and the melt, $\tau_q$ (bottom), versus $T - \Tc$ in a double logarithmic plot for selected $q$-values as indicated. Symbol sizes are one standard deviation. In this plot, $\Tc = 0.450$ is kept fixed. According to equation~(\ref{eq:ttsp_tau}), the plot should yield straight lines with a slope $\gamma=2.09$. Furthermore, MCT expects the power law to extend to larger temperatures than those, where the $\beta$-process can be observed \cite{FranoschFuchsGoetze1997}.  In accord with the latter expectation, the simulation data exhibit a power-law behaviour for $T \leq 0.7$ (except for the largest $q$-values), whereas a quantitative analysis of the $\beta$-process was only possible for $T\leq 0.52$ (see part~I and 
\cite{BennemannBaschnagelPaul1999_incoherent}).  However, contrary to the MCT prediction, there is a non-negligible variation of $\gamma$  with $q$. The solid lines show least-square fits, including the maximum number of temperatures, to determine the exponent $\gamma$. The dashed line shows $(T-\Tc)^{-2.09}$ for comparison.  
}
\efig

The region where the relaxation times exhibit a linear behaviour in Figure~\ref{fig:log_tau_alpha_chain+melt_vs_log_T-T_c} were used to calculate the exponent $\gamma$ with least square fits.  The same analysis had already been done for the incoherent scattering function in \cite{BennemannPaulBaschnagel1999}.  Figure~\ref{fig:gamma_inc_chain_melt} displays the combined results as a function of the wave numbers $q$. Qualitatively, the same behaviour is observed for all correlators. There is a systematic drift of $\gamma$ with $q$, which is not accounted for by (the leading-order predictions of) ideal MCT.  The outcome of the $\beta$-analysis \cite{BennemannBaschnagelPaul1999_incoherent}, $\gamma = 2.09$, only represents a good approximation for $3 \lesssim q \lesssim 8$.  For large $q$ it underestimates the adjusted $\gamma$-values, whereas it overestimates them when $q \rightarrow 0$. In this low-$q$ limit the simulation results tend towards the exponent $\gamma_D$ determined from the diffusion coefficient of a chain ($D \sim (T-\Tc)^{\gamma_D}$), i.e., $\gamma_{D} = 1.84 \pm 0.02$.

To cross-check the results of Figure~\ref{fig:log_tau_alpha_chain+melt_vs_log_T-T_c} we also plotted $(\tau^\mr{x}_q)^{-1/\gamma}$ versus $T$ with fixed $\gamma = 2.09$.  From equation~(\ref{eq:ttsp_tau}) it is expected to find straight lines for each $q$, which intersect the temperature axis at $\Tc$.  As in Figure~\ref{fig:log_tau_alpha_chain+melt_vs_log_T-T_c} visible deviations from a power law occur for all $q$ at $T=0.46$ and at large $q$ for the highest temperatures.  Applying least square fits to the data points for $3 \leq q \leq 16$ in the temperature interval $0.47 \leq T \leq 0.52$, for which the $\beta$-analysis was possible (see \cite{betaDynamics}), we found a systematic shift of $\Tc$ to smaller values with decreasing $q$.  Qualitatively, this finding resembles that of $\gamma$.  However, the $q$-dependence of the critical temperature is weaker than that of $\gamma$ so that the results are compatible with $\Tc = 0.450 \pm 0.005$ within the error bars if $q \gtrsim 3$.  This agrees with the analysis of the incoherent scattering function \cite{BennemannPaulBaschnagel1999,BennemannBaschnagelPaul1999_incoherent}.
 
\bfig
\rsfig{figures/gamma_inc_chain_melt.eps} 
\caption[]{
\label{fig:gamma_inc_chain_melt}
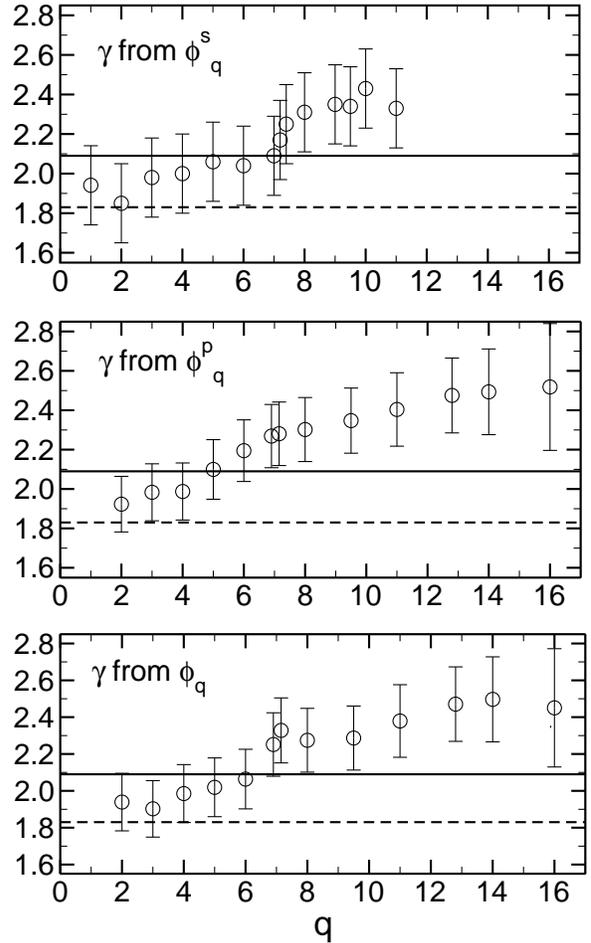
$\alpha$-relaxation exponents $\gamma$ from the incoherent scattering function (top, from \cite{BennemannBaschnagelPaul1999_incoherent}) and the coherent counterparts versus $q$ extracted by least-square fits (see Figure~\ref{fig:log_tau_alpha_chain+melt_vs_log_T-T_c}).  $\gamma = 2.09 (\pm 0.07)$, the result from the $\beta$-analysis \protect \cite{BennemannBaschnagelPaul1999_incoherent}, is shown as a thick solid line.  To indicate the $q \rightarrow 0$ limit the exponent, $\gamma_{D} = 1.84 (\pm 0.02)$, from an analysis of the diffusion coefficient of a chain is drawn as a thick dashed horizontal line.
}
\efig

\subsection{Time-Temperature Superposition Principle}
\label{subsec:ttsp}

\bfig
\rsfig{figures/coh_melt_sf_rescaled_q3.0_allT.eps}
\rsfig{figures/coh_melt_sf_rescaled_q6.90_allT.eps}
\rsfig{figures/coh_melt_sf_rescaled_q14.0_allT.eps}
\caption[]{
\label{fig:coh_melt_sf_rescaled_q3.0_q6.9_q14.0_allT}
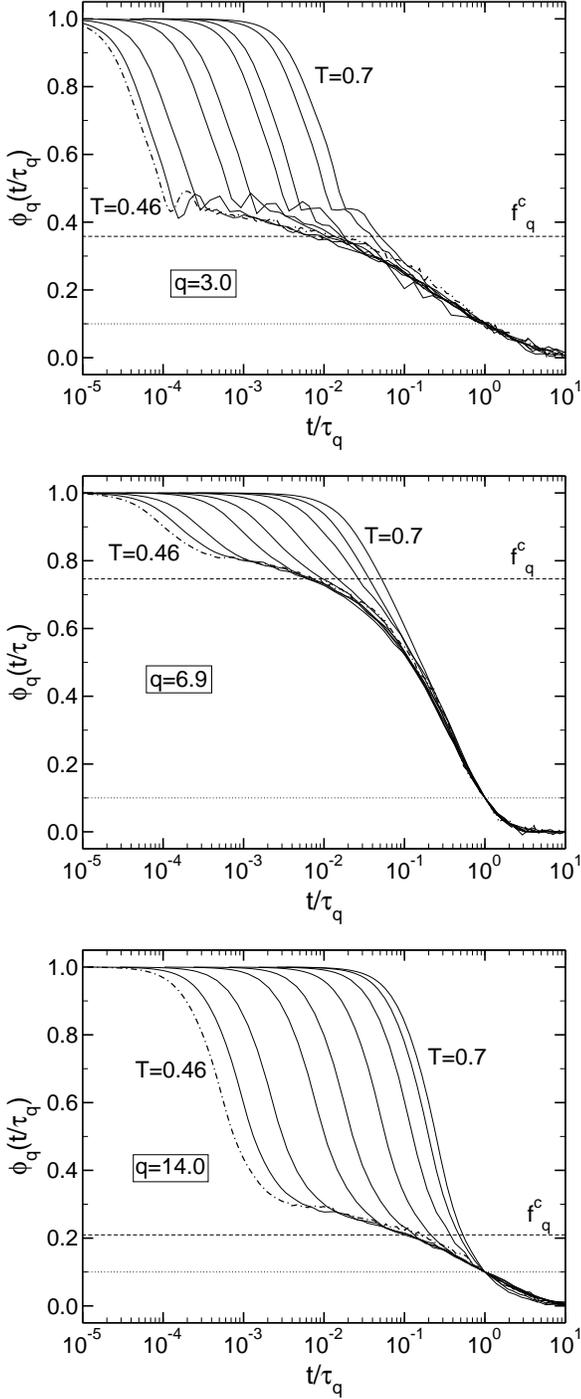
Demonstration of the time-temperature superposition principle (TTSP) for the coherent intermediate scattering function $\phi_q(t)$ for all investigated temperatures, $T=0.46$, 0.47, 0.48, 0.50, 0.52, 0.55, 0.6, 0.65, 0.7 (from left to right in the plots) for $q=3.0$ (top), $q=6.9$ (middle), and $q=14.0$ (bottom).  At $T=0.46$ (dot-dashed line) the TTSP does not work as well as for the other temperatures.  The $\alpha$-relaxation times were taken to be the decay time to 0.1 (dotted horizontal line).  For each $q$ the non-ergodicity parameters $f^\mr{c}_q$ are shown as dashed lines.
}
\efig

Figure~\ref{fig:coh_melt_sf_rescaled_q3.0_q6.9_q14.0_allT} tests the validity of the time-temperature superposition principle (TTSP) expressed by equation~(\ref{eq:ttsp_phi}).  It shows the coherent scattering functions versus time rescaled by the respective $\alpha$-relaxation times, i.e.\ $\phi_q(t/\tau_q)$, for $q=3.0$, 6.9, and 14.0 at all investigated temperatures.  The relaxation times are defined by $\phi_q^\mr{x}(\tau^\mr{x}_q) = 0.1$.  For this choice all correlators, even at high $q$, are sufficiently in the $\alpha$-regime so that the $\beta$-process does not perturb the behaviour.  In the late $\alpha$-regime the curves collapse onto a master curve, with visible deviations for $q=14.0$ at the highest temperatures.  Figure~\ref{fig:log_tau_alpha_chain+melt_vs_log_T-T_c} already showed that the power law for the relaxation times is violated for $q=16$ at $T \geq 0.65$.  Thus, it is not surprising to find deviations from the TTSP master curves for similar $q$ and $T$ in Figure~\ref{fig:coh_melt_sf_rescaled_q3.0_q6.9_q14.0_allT}, too.  

With decreasing temperature the TTSP property starts at earlier rescaled times and extends well above $f_q^\mr{c}$, a result often obtained from computer simulations \cite{BennemannPaulBinder1998,KaemmererKobSchilling1998,KobAndersen_LJ_II_1995} and numerical calculations \cite{FuchsHofackerLatz1992}.  This trend is violated for the lowest simulated temperature $T=0.46$: Especially in the early $\alpha$-region the curves for $T=0.46$ deviate from the master curve.  We believe that this observation is not a non-equilibrium effect as we compared $\phi_q(t)$ from 10 simulation runs shifted in time by about $10 \times \tau_{q_\mr{max}}$ and found no significant differences.  An explanation could rather be that equation~(\ref{eq:ttsp_phi}) is no longer valid because $T=0.46$ is so close to $\Tc$ that relaxation processes, which are not taken into account by idealized MCT, begin to dominate the dynamics.  For the $\beta$-relaxation the influence of these additional processes (called ``hopping processes'' by MCT) have been worked out in the extended mode-coupling theory \cite{FuchsGoetzeHildebrand1992_extMCT}.  This analysis shows that these relaxation processes change the shape of $\phi^\mr{x}_q(t)$ in the late-$\beta$/early-$\alpha$ regime and remove the power-law divergence of the $\alpha$-time scale $\tilde{\tau}$ (see Eq.~(\ref{eq:tau_alpha})).  Therefore, it is possible that deviations from the idealized TTSP occur if $T$ is very close to $\Tc$.  Such deviations have already been observed for $\phi^\mr{s}_q(t)$ for the present model \cite{BennemannBaschnagelPaul1999_incoherent}.    

\bfig
\rsfig{figures/coh_melt_sf_low_q_T0.47.eps}
\caption[]{
\label{fig:coh_melt_sf_low_q_T0.47}
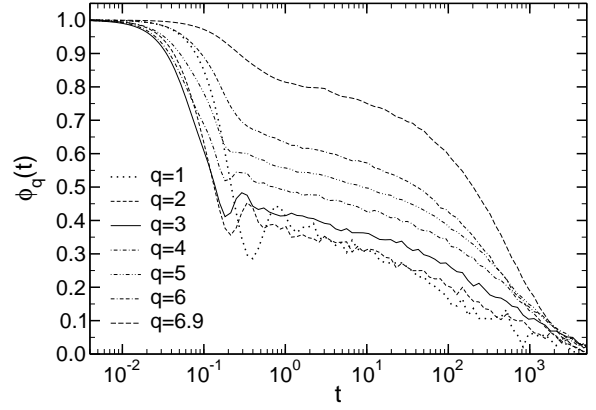
Coherent intermediate scattering function $\phi_q(t)$ at $T=0.47$ for the lowest investigated wave-numbers $q$ up to $q_\mr{max}=6.9$.  At small $q$ an oscillation at $0.15 \lesssim t \lesssim 1.0$ develops, which is seen neither for $\phi^\mr{s}_q(t)$ \protect \cite{BennemannBaschnagelPaul1999_incoherent} nor for $\phi^\mr{p}_q(t)$.
}
\efig

For $q \lesssim 4$ an oscillation at the beginning of the $\beta$-regime can be observed at all temperatures, which becomes more pronounced at lower $T$.  With decreasing temperature the amplitude of the oscillation increases when $q$ becomes smaller, as exemplified in Figure~\ref{fig:coh_melt_sf_low_q_T0.47} for $T=0.47$.  Sound waves travelling through the simulation box and reentering again at the opposite side could be the reason for this effect \cite{LewisWahnstrom1994}.  The sound velocity was estimated to be $c \approx 7$ \cite{BennemannBaschnagelPaul1999_incoherent}.  Using $2\pi/c q = t^\mr{osc}_q$ we obtain times $t^\mr{osc}_q$ comparable with the times where the oscillations occur in Figure~\ref{fig:coh_melt_sf_low_q_T0.47}.  Note that neither $\phi^\mr{s}_q(t)$ \cite{BennemannBaschnagelPaul1999_incoherent} nor $\phi^\mr{p}_q(t)$ show such oscillations.  This might be a hint that a non-linear coupling between different $q$-values of the coherent scattering function is important for their presence. 

Another feature of the data are weak bumps at the beginning of the $\beta$-plateau, which occur at $t \approx 3$ for all correlators and all $q$ if $T \leq 0.48$.  These are discussed in \cite{BennemannBaschnagelPaul1999_incoherent} for $\phi^\mr{s}_q(t)$.  They are hard to see in Figures~\ref{fig:coh_melt_sf_rescaled_q3.0_q6.9_q14.0_allT} and \ref{fig:coh_melt_sf_low_q_T0.47} for $T=0.47$ because the statistics is much worse for $\phi_q(t)$ than for $\phi^\mr{s}_q(t)$. So, it is difficult to distuinguish them from statistical fluctuations.  In part I a weak bump is also visible in Figures~4 and 12 at $t \approx 2$.  As the aforementioned oscillations these bumps could be related to sound waves \cite{LewisWahnstrom1994}.  Using $L=10.5$ for the linear dimension of the simulation box and $c \approx 7$, one obtains $t = L/c \approx 1.5$, which is close to the position of the bumps.  If they were really related to sound waves, their position should shift when changing the size of the simulation box.  We have not attempted such a finite size study yet, although simulation results on two-dimensional polymer melts \cite{RayBinder1994}, on binary mixtures of LJ-particles \cite{BuechnerHeuer_PRE1999} or of hard spheres \cite{KimYamamoto2000} and on silica glasses \cite{Horbach_PRE1996,Horbach_PhilMag1999} suggest that there might be pronounced effects at low temperature. 

Besides sound waves the oscillations could still have other causes, for instance, non-equilibrium effects \cite{KobBarrat2000}.  However, this cause seems unlikely, since the oscillations are equally present at all investigated temperatures and great care was taken to achieve equilibrium at the lowest temperatures.  Furthermore, the Nos{\'e}-Hoover thermostat used to maintain constant temperature in the simulation can also be ruled out as a crucial factor.  In Ref.~\cite{BennemannPaulBinder1998} its influence was thoroughly tested against simulations in the microcanonical ensemble.  No significant difference of the dynamics was found. 

Of course, it is important to assess the relevance of the oscillations for the results in the $\alpha$-regime.  The analysis of the $\beta$-process in part I showed that the oscillations perturb the $\beta$-master curve only at the beginning of the $\beta$-regime for $q \lesssim 3$ (see Figure~5 of \cite{betaDynamics}).  For longer times no deviations can be observed in comparison to the behaviour of the correlators at larger $q$-values, where the oscillations are absent.  If the major part of the $\beta$-process is already unaffected, it is unlikely that the oscillations could have a pronounced influence on the $\alpha$-relaxation discussed here.

\subsection{Kohlrausch-Williams-Watts Fits}
\label{subsec:kww}
When performing KWW-fits one has to keep in mind that the results will vary quite significantly with the chosen fit interval  (see below and also \cite{FuchsHofackerLatz1992,CDFGHLLTerwMCT1993}).  Since the $\beta$-process overlaps with the $\alpha$-process, one would have to use a combination of equations~(\ref{eq:early_alpha}) and (\ref{eq:KWW-function}) in order to incorporate the influence of the $\beta$-dynamics.  To alleviate this problem sometimes only the late $\alpha$-process is fitted \cite{KaemmererKobSchilling1998,KobAndersen_LJ_II_1995} to minimize interference with the $\beta$-process (see also \cite{BennemannBaschnagelPaul1999_incoherent} for a more detailed discussion of this and other approaches).  Apart from these fundamental considerations it is clear that a highly non-linear fit will sensitively depend on its input data.  

We fitted KWW-functions using two different procedures: One method was a fit with three adjustable parameters ($f^\mr{x K}_q$, $\tau^\mr{x K}_q$, $\beta^\mr{x K}_q$).  This produces a description which is not biased by any theoretically motivated restrictions on (some of) the fit parameters.  In the other method $f^\mr{x K}_q$ was taken to be the non-ergodicity parameter (NEP) $f^\mr{xc}_q$.  This choice is suggested by the MCT prediction (\ref{eq:early_alpha}).  It also fixes the KWW-times via $\phi^\mr{x}_q(\tau^\mr{x K}_q) = f^\mr{xc}_q /\mr{e}$ so that the stretching exponent $\beta^\mr{x K}_q$ is the only free parameter.  In the following the results of both approaches, which we call ``free fit'' and ``fixed-NEP fit'', are compared.

The constraint $f^\mr{x K}_q = f^\mr{xc}_q$ is not a very strong restriction: the fitted values for $f^\mr{x K}_q$ are within 10\% of the non-ergodicity parameter $f^\mr{xc}_q$.  The deviations of $f^\mr{x K}_q$ relative to $f^\mr{xc}_q$ depend on both $q$ and the size of the fit interval ($= x_\mr{cut}$, see below) .  If $q > q_\mr{max}$, $f^\mr{x K}_q$ is almost independent of the chosen fit interval and systematically larger than $f^\mr{xc}_q$.  For $q < q_\mr{max}$ the influence of the fit interval becomes bigger. Furthermore, $f^\mr{x K}_q < f^\mr{xc}_q$ becomes possible.  Despite these (quantitative) differences both $f^\mr{x K}_q$ and $f^\mr{xc}_q$ exhibit the same qualitative $q$-dependence.  Consequently, the KWW-times from free and fixed-NEP fits also deviate only quantitatively (by at most 15\% for $q \geq 4$), but not qualitatively.  Both fit procedures yield the same $q$-dependence for the relaxation time.  We will return to this point below.

For the fits all data points with $\phi^\mr{x}_q(t) < x_\mr{cut} f^\mr{xc}_q$ were used.  The fit interval was systematically varied by increasing the cutoff parameter $x_\mr{cut}$ from 0.3 to 0.9 in steps of 0.1 to investigate the influence of the chosen interval on the resulting KWW-parameters.  Removing data below a certain small positive value leads to systematically higher stretching exponents, especially at large $q$.  Therefore, we did not introduce a lower bound, although noise around zero might influence the results.  However, such a lower bound -- data points below 0.02 were discarded -- was used in the study of $\phi_q^\mr{s}(t)$ in \cite{BennemannBaschnagelPaul1999_incoherent}, resulting in greater values for $\beta^\mr{s}_q$ than in our analysis.  Nevertheless, the difference is less than 5\% at the largest $q$. So, the statements made in \cite{BennemannBaschnagelPaul1999_incoherent} remain valid.  Because of the TTSP the shape of $\alpha$-relaxation is identical for all temperatures $0.47 \leq T \leq 0.52$ (cf.\ Figure~\ref{fig:coh_melt_sf_rescaled_q3.0_q6.9_q14.0_allT}).  Thus, we confined the KWW-analysis to $T=0.47$.

\bfig
\rsfig{figures/beta_K_coh_melt_free+fixed_NEP+S_q.eps}
\caption[]{
\label{fig:beta_K_coh_melt_free+fixed_NEP+S_q}
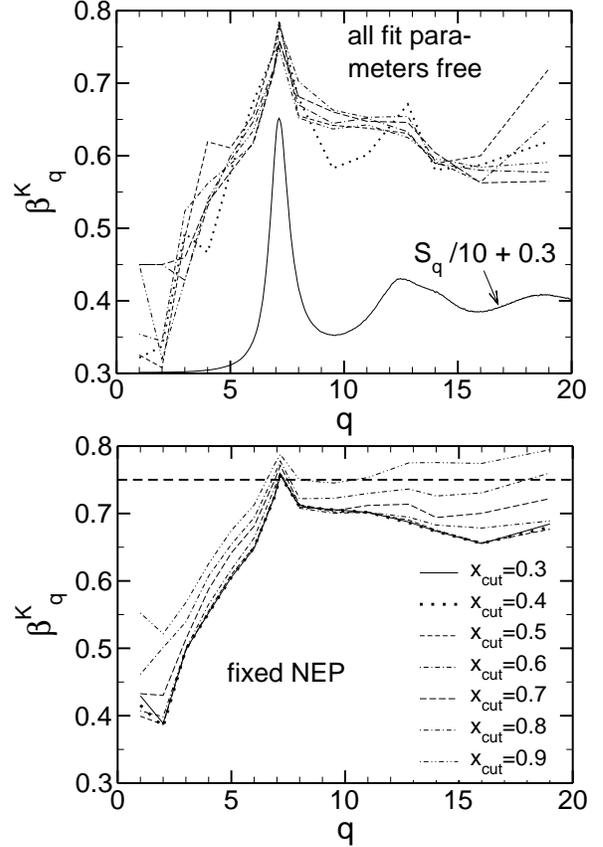
KWW-exponent $\beta^\mr{K}_q$ of the coherent scattering function $\phi_q(t)$ at $T=0.47$ for different cutoff values $x_\mr{cut} f^\mr{c}_q$.  For a given $x_\mr{cut}$ all data points with $\phi_q < x_\mr{cut} f^\mr{c}_q$ were used for the fits.  A thick dashed line is drawn at $b = 0.75$, the prediction of MCT for $q \rightarrow \infty$ (Eq.~\ref{eq:limit_beta})).  In the upper panel all three KWW parameters were fitted.  In the lower panel only $\beta^\mr{K}_q$ was adjusted. $f^\mr{K}_q = f^\mr{c}_q$ was fixed and thus also the time constant $\tau^\mr{K}_q$ by $\phi_q(\tau^\mr{K}_q) = f^\mr{c}_q/\mr{e}$.  One sees that $\beta^\mr{K}_q$ is qualitatively in phase with the structure factor $S_q$, especially when all three free parameters are allowed to vary freely.  The line types for both panels are identical.  In the upper one the curve for $x_\mr{cut}=0.3$ is missing because the fit algorithm did not converge. 
}
\efig

A comparison of the results for the stretching exponent $\beta^\mr{K}_q$ obtained from fits with three free parameters (top) and with fixed $f^\mr{K}_q = f^\mr{c}_q$ (bottom) is shown in Figure~\ref{fig:beta_K_coh_melt_free+fixed_NEP+S_q}.  Qualitatively, both plots are comparable.  They show that $\beta^\mr{K}_q$ is roughly in phase with the static structure factor, a result also found in calculations for hard sphere systems \cite{FuchsHofackerLatz1992}, simulations of water \cite{SciortinoFabbianChen1997} or neutron scattering experiments of orthoterphenyl \cite{ToelleWuttkeSchober1998}.  On the other hand, the data indicate a further feature which is not present in $S_q$.  There is a weak shoulder at $q \approx 4.5 \approx 2\pi/R_\mr{g}$.  This seems to correspond to the large $\tau_q$ encountered in Figure~\ref{fig:tau_q_and_S_q_vs_q_T046_T052_T07}.

Let us now compare both fit methods in more detail.  The resulting values of $\beta^\mr{K}_q$ differ especially at small and large $q$.  At small $q$, this is not surprising, since the differences between fitted $f^\mr{x K}_q$ and $f^\mr{xc}_q$ are most pronounced in this $q$-regime.  In the case of the fixed-NEP fit, $\beta^\mr{K}_q$ increases monotonously with the cutoff parameter when $x_\mr{cut} \geq 0.6$, whereas it does not significantly depend on it for $x_\mr{cut} \leq 0.5$.  This is an indication that the presence of the $\beta$-process starts to be felt at $x_\mr{cut} \geq 0.6$.  As the other two parameters are fixed, a better adaption to the $\beta$-decay can only be accomplished by an increasing stretching exponent (note that $b=0.75 > \beta^\mr{K}_q$ for these small $q$-values).  Contrary to that, the fits with three free parameters do not show a systematic dependence on $x_\mr{cut}$.  The curves scatter randomly by roughly $15\%$.  The size of this variation is comparable to that of the fixed-NEP fit.  Nevertheless, the sensitivity to $x_\mr{cut}$ and the different results of both methods illustrate that the choice of the fit conditions can lead to (significantly) different results, although the statistical accuracy of the input data is good (see Fig.~\ref{fig:KWW_all_fixed+all_free_q_7.15_8.0_9.5_16.0+late_alpha_zoom}).

As $q$ approaches $q_\mr{max}$, the $\beta^\mr{K}_q$-values of both methods converge and essentially agree with each other at $q_\mr{max}$, where stretching is least pronounced.  However, if $q$ increases beyond $q_\mr{max}$, the results of both methods differ again substantially.  The fixed-NEP fit exhibits a strong dependence on $x_\mr{cut}$, which is quite generally much larger than that of the free fit, and gives results close to the large-$q$ limit $\beta_q^\mr{K}=b$ (see Eq.~(\ref{eq:limit_beta})), whereas the $\beta^\mr{K}_q$-values of the free fit are distinctly smaller and do not seem to tend to $b$.  We only see an increase of $\beta^\mr{K}_q$ at the largest $q$.  This might suggest either that the previously determined exponent parameter $\lambda$ is too small or that the predicted large-$q$ limit is approached from below -- if at all -- in our system.  The latter behavior would be in contrast to calculations for hard spheres \cite{FuchsHofackerLatz1992} and simulations of water \cite{SciortinoFabbianChen1997,StarrSciortino1999}.  A test of this conjecture would require an extension of the analysis to greater $q$.  However, this extension is hampered by the very small plateau values at large $q$, which renders it difficult to discriminate the $\alpha$-decay from the statistical noise around zero.

In calculations for ideal hard spheres the limit of $\beta^\mr{K}_q = b$ is reached at $q \approx 6 \times q_\mr{max}$ \cite{Fuchs1994_kww}.  For our system this would correspond to $q \approx 42$.  This might be the reason why $\beta^\mr{K}_q$ does not approach the large-$q$ limit $b$ in the range of wave vectors investigated for most fit intervals in the upper panel of Figure~\ref{fig:beta_K_coh_melt_free+fixed_NEP+S_q}. Similarly, a KWW-analysis of the incoherent scattering of a binary Lennard-Jones mixture could not confirm the large-$q$ behaviour of $\beta^\mr{s K}_q$ by means of a free fit procedure \cite{KobAndersen_LJ_II_1995}.  However, the lower panel of Figure~\ref{fig:beta_K_coh_melt_free+fixed_NEP+S_q} suggests that the choice $\beta^\mr{K}_q = b$ is not unreasonable if $q > q_\mr{max}$.  This conclusion was also drawn from a fixed-NEP analysis of incoherent scattering in \cite{BennemannBaschnagelPaul1999_incoherent}.  Therefore, we performed the comparison of both methods again for $\phi_q^\mr{s}(t)$ and also for coherent chain scattering.  The results for $\beta^\mr{s K}_q$ and $\beta^\mr{p K}_q$ are the same as in the case of $\beta_q^\mr{K}$. The behavior observed in Figure~\ref{fig:beta_K_coh_melt_free+fixed_NEP+S_q} is thus representative of all scattering functions of our model. 

\bfig
\rsfig{figures/KWW_all_fixed+all_free_q_7.15_8.0_9.5_16.0+late_alpha_zoom.eps}
\caption[]{
\label{fig:KWW_all_fixed+all_free_q_7.15_8.0_9.5_16.0+late_alpha_zoom}
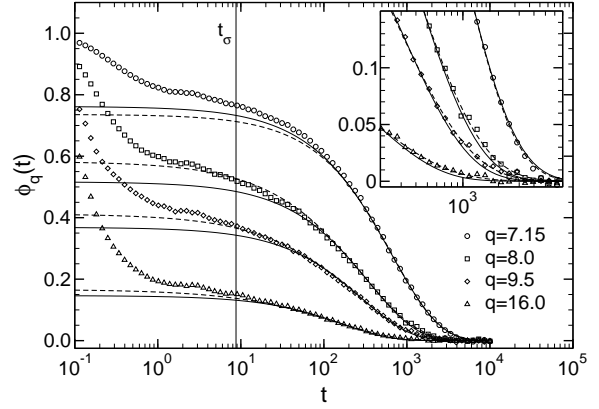
Comparison between $\phi_q(t)$ from simulation data (symbols), KWW-fits with three free parameters to data points with $\phi_q(t) < f^\mr{c}_q/2$ (dashed lines) and KWW-functions calculated from the large-$q$ prediction of MCT: $\lim_{q \rightarrow \infty} \beta^\mr{K}_q =b$ and $\phi(\tau^\mr{K}_q) = f^\mr{c}_q /\mr{e}$ (solid lines) at some $q \geq q_\mr{max}$ as indicated.  The temperature is $T=0.47$.  $t_\sigma$ is the $\beta$-relaxation time \protect \cite{BennemannBaschnagelPaul1999_incoherent}.  The figure shows that the large-$q$ approximation (\ref{eq:kww_limit}) works quite well, but from the inset it can be seen that it does not describe the late $\alpha$-process as well as the free fit.
}
\efig

Therefore the question arises of whether a KWW-func\-tion with $f^\mr{K}_q = f^\mr{c}_q$ and $\beta^\mr{K}_q = b$ can provide a comparable description of the simulation data as a free fit at large $q$.  Figure~\ref{fig:KWW_all_fixed+all_free_q_7.15_8.0_9.5_16.0+late_alpha_zoom} shows a comparison of both approaches for $\phi_q(t)$ at various $q > q_\mr{max}$.  At first sight, both methods yield very satisfactory approximations for the $\alpha$-decay, but $\chi^2$ is typically a factor of 10 smaller for the free fit.  The origin of this difference is illustrated in the inset of the figure: Using $f^\mr{K}_q = f^\mr{c}_q$, $\beta^\mr{K}_q = b$, and $\tau^\mr{K}_q$ from $\phi^\mr{x}_q(\tau^\mr{K}_q) = f^\mr{c}_q /\mr{e}$, the description of the data works quite well for $q \gtrsim q_\mr{max}$, except for the very late $\alpha$-process which are better described by the free fit.  The discrepancy does not diminish if $\beta^\mr{x K}_q$ is adjusted instead of posing $\beta^\mr{K}_q = b$ while still keeping $f_q^\mr{K} = f_q^\mr{c}$.  The same results are also obtained for the incoherent scattering and the coherent chain scattering functions.  So, we may conclude that the free fit procedure yields a better description of the simulation data than the fixed-NEP fit and that we cannot unambiguously confirm the asymptotic limit $\beta_q^\mr{x K}=b$ in the $q$-range accessible to our simulation.

\bfig
\rsfig{figures/tau_kww_inc+chain+melt_T0.47_vs_q.eps}
\caption[]{
\label{fig:tau_kww_inc+chain+melt_T0.47_vs_q}
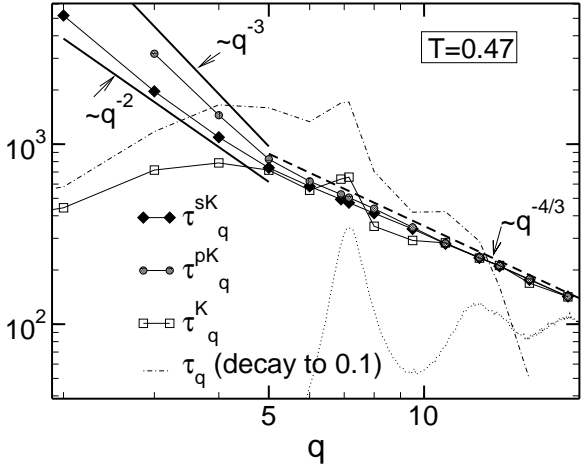
KWW-times $\tau^\mr{x K}_q$ (symbols) defined by $\phi^\mr{x}_q(\tau^\mr{x K}_q) = f^\mr{xc}_q /\mr{e}$ from simulation data versus $q$ in a double logarithmic plot. The $q$-values shown are: $q=2$, 3, 4, 5, 6, 6.9, 7.15, 8, 9.5, 11, 12.8, 14, 16, 19. $\tau^\mr{s K}_q$ and $\tau^\mr{p K}_q$ are accurate to within 5\%, $\tau^\mr{K}_q$ to within 10\% (two $\sigma$ intervals).  The thick dashed line is the large-$q$ prediction of MCT, $\tau^\mr{x K}_q \propto q^{-1/b}$ (Eq.~(\ref{eq:limit_tau_kww})), which is seen to be nicely fulfilled for all correlators.  $S_q$ (scaled and shifted) is indicated with a dotted line.  The $\alpha$-decay times $\tau_q$ (defined by $\phi_q(\tau_q) = 0.1$) are also shown as a dash-dotted line.  Except at large $q$ ($q \gtrsim 12.8$) they exhibit the same qualitative behaviour as the KWW-times $\tau^\mr{K}_q$.  Power laws $q^{-2}$ and $q^{-3}$ are shown for comparison.
}
\efig

However, the situation is different for the KWW-times.  These times can be defined by $\phi^\mr{x}_q(\tau^\mr{x K}_q) = f^\mr{xc}_q /\mr{e}$ (or by $\phi^\mr{x}_q(\tau^\mr{x K}_q) = f^\mr{x K}_q /\mr{e}$), independently of $\beta_q^\mr{K}$. Since the $q$-dependence of $f^\mr{xc}_q$ closely matches that of $f^\mr{x K}_q$ and both KWW-prefactors are numerically not very different, the KWW-times from the free fits show the same behaviour as those of the fixed-NEP fits. So, only the KWW-times, defined by $\phi^\mr{x}_q(\tau^\mr{x K}_q) = f^\mr{xc}_q /\mr{e}$, are shown for $T=0.47$ in Figure~\ref{fig:tau_kww_inc+chain+melt_T0.47_vs_q} and are compared with the collective $\alpha$-relaxation times $\tau_q$.  If $q \lesssim 12.8$, $\tau_q$ and $\tau^\mr{K}_q$ exhibit a qualitatively similar behaviour: There is a broad maximum around $q \approx 4.5$ and a sharp maximum at $q_\mr{max}$ followed by oscillations which are in phase with each other and mimic those of $S(q)$.  However, the absolute values of both relaxation times differ.  If $q \lesssim 12.8$, $\tau_q > \tau_q^\mr{K}$, whereas $\tau_q < \tau_q^\mr{K}$ for larger $q$.  At these large $q$-values the behaviour of both relaxation times is also qualitatively different.  $\tau_q$ decreases sharply, whereas  all $\tau^\mr{x K}_q$ curves fall onto a master curve if $q \gtrsim 11$.  This master curve is given by a power law which is very close to $q^{-4/3}$, the asymptotic result of MCT (see Eq.~(\ref{eq:limit_tau_kww})).

This disparity between $\tau^\mr{x}_q$ and $\tau_q^\mr{x K}$ can be rationalized by the following argument.  According to equation~(\ref{eq:ttsp_tau}) the times
\begin{equation}
\label{eq:tau_explicit}
\tau^\mr{x}_q = C^\mr{x}(q, d=0.1) \tilde{\tau}; \quad \tau^\mr{x K}_q = C^\mr{x}(q, f^\mr{xc}_q/\mr{e}) \tilde{\tau} \;, 
\end{equation}
are related to one another.  Assuming a KWW-behaviour with $f^\mr{x K}_q=f^\mr{xc}_q$ for the correlators in the $\alpha$-regime, we get
\begin{equation}
\label{eq:tau_kww_and_alpha}
\frac{\tau^\mr{x}_q}{\tau^\mr{x K}_q}  = \frac{C^\mr{x}(q, d)}{C^\mr{x}(q, f^\mr{xc}_q/\mr{e})} = \left[\ln(f^\mr{xc}_q/d)\right]^{1/\beta^\mr{x K}_q}, \quad f^\mr{xc}_q > d \;.
\end{equation}
Above relation was tested for the coherent (chain and melt) $\alpha$-decay and KWW-times.  In the range $3 \leq q \leq 14$ and for all $x_\mr{cut}$ the deviation is smaller than 10\%.  This is a strong consistency test because the $\alpha$-times $\tau^\mr{x}_q$ were determined completely independently of the KWW-times, in which the non-ergodicity parameters from the $\beta$-analysis also enter.  If $q \lesssim 12.8$, the choice $d=0.1$ implies that $f^\mr{xc}_q/d > \mr{e}$ so that $\tau_q > \tau_q^\mr{K}$, and vice versa if $q > 12.8$.  Furthermore, the previous discussion revealed that $f^\mr{c}_q$ and $\beta^\mr{K}_q$ are in phase with $S_q$.  By evaluating the right-hand side of Eq.(\ref{eq:tau_kww_and_alpha}) one sees that the ratio $\tau_q/\tau^\mr{K}_q$ is modulated with $S_q$ as well.  This modulation should be more pronounced for $\tau_q$ than for the KWW-times $\tau^\mr{K}_q$ because the in-phase variation of $\tau^\mr{K}_q$ with $S_q$ should amplify that of the right-hand side of equation~(\ref{eq:tau_kww_and_alpha}).  Figure~\ref{fig:tau_kww_inc+chain+melt_T0.47_vs_q} supports this expectation if $q \lesssim 12.8$.  On the other hand, if $q > 12.8$, the $q$-dependence of the right-hand side does no longer strongly oscillate, but is rather close to a power law which modifies that of $\tau_q^\mr{x K}$.  This comparison shows that the asymptotic behavior of equation~(\ref{eq:tau_kww_and_alpha}) can only be observed if the $q$-dependence of the amplitude of the $\alpha$-process is taken into account, as it was also done in the theoretical derivation of equation~(\ref{eq:tau_kww_and_alpha}) \cite{Fuchs1994_kww}.

At this point, we would like to note that, to the best of our knowledge, neither a modulation of $\tau^\mr{K}_q$ with $S_q$ nor a prepeak at wave vectors corresponding to the radius of gyration has been reported in neutron scattering experiments.  For instance, experiments on polybutadiene \cite{ArbeRichterColmenero1996} or polyisobutylene \cite{RichterMonkenbuschAllgeier1999} rather find evidence for a power law $q^{-n}$ ($n = 3.1$ \cite{RichterMonkenbuschAllgeier1999}, $n = 3.6$ \cite{ArbeRichterColmenero1996}) in the range of the first and second maxima of $S_q$.  The power law is superimposed with oscillations that seem to in phase with $S_q$ for polyisobutylene \cite{RichterMonkenbuschAllgeier1999,RichterMonkenbuschArbe1999}, but in anti-phase for polybutadiene \cite{ArbeRichterColmenero1996}.  A possible origin of this difference between experiment and simulation could be the temperature range studied.  In the simulation, the modulation of $\tau_q$ (or $\tau^\mr{K}_q$) with the coherent static structure factor only becomes visible very close to $\Tc$.  $T=0.47$ corresponds to a reduced temperature of $(T-\Tc)/\Tc \approx 0.044$, whereas the measurements of $S_q$ of polybutadiene \cite{ArbeRichterColmenero1996} were done at 260 K, so at a reduced temperature of $\approx 0.2$ (using $T^\mr{PB}_\mr{c} \approx 216\, \mr{K}$ \cite{ZornRichterFrick1993,FrickFaragoRichter1990}).  In turn, this reduced temperature would correspond to $T=0.54$ in our model, a temperature at which a modulation with $S_q$ is almost not present (see Figure~\ref{fig:tau_q_and_S_q_vs_q_T046_T052_T07}).  The same conclusion should also hold for polyisobutylene because the temperatures investigated, $T = 390,470 \, \mr{K}$, are far above $T_\mr{g} = 201 \,\mr{K}$ \cite{BoehmerNgaiAngellPlazek1993}.  Furthermore, the prepeak on the scale of the radius of gyration has perhaps not been found in the experiments, since it would lie outside of the $q$-range studied in the coherent scattering of the melt.  An estimate of the corresponding $q$-value with the experimental value for $R_\mr{g}$ ($\simeq 19$ {\AA}) for polyisobutylene suggests this conjecture.

At small $q$, $\tau^\mr{s K}_q$ increases less steeply than $\tau^\mr{p K}_q$. This might be rationalized by taking into account that the chain has a structure, which decays in time, whereas no such additional relaxation happens for the single particle correlator.  At the smallest $q$-values ($q \leq 5$) $\tau^\mr{p K}_q$ follows approximately a $q^{-3}$ power law behaviour, which is comparable to experimental results \cite{RichterMonkenbuschAllgeier1999}.  Free diffusion would require a $q^{-2}$ behaviour, expected for $q \ll 1 / R_\mr{g} \approx 0.7$ \cite{DoiEdwards}, which is not accessible in our investigation.  At the lowest $q$-values, for which KWW-fitting was possible, $\tau^\mr{s K}_q$ increases more steeply than $q^{-2}$ in agreement with previous findings \cite{BennemannBaschnagelPaul1999_incoherent}.  Note that the investigated relaxation times in \cite{BennemannBaschnagelPaul1999_incoherent} were defined by a decay to $d=f^\mr{sc}_q /2$ instead of $d=f^\mr{sc}_q /\mr{e}$ used here.  According to equation~(\ref{eq:tau_kww_and_alpha}), this implies that the two times differ by a factor of $C^\mr{s}(q, f^\mr{sc}_q/2) / C^\mr{s}(q, f^\mr{sc}_q/\mr{e}) \approx (\ln 2 )^{1/\beta^\mr{s K}_q}$. This function varies slowly with $q$ so that the $q$-dependence of both relaxation times should only be slightly different. 

Finally at $q \approx 6$, $\tau^\mr{s K}_q$ and $\tau^\mr{p K}_q$ cross over to the large-$q$ prediction $\tau^\mr{x K}_q \propto q^{-1/b}$ (Eq.~(\ref{eq:limit_tau_kww})) of MCT.  Due to the modulation of $\tau^\mr{K}_q$ with $S_q$ the collective relaxation time of the melt enters the asymptotic regime later than $\tau^\mr{s K}_q$ or $\tau^\mr{p K}_q$ (for $q \gtrsim 11$).  It is interesting to note that the large-$q$ behaviour of MCT is seen rather clearly for all $\tau^\mr{x K}_q$, whereas the large-$q$ behaviour of $\beta^\mr{x K}_q$ could not be confirmed unambiguously -- if at all (cf.\ Figure~\ref{fig:beta_K_coh_melt_free+fixed_NEP+S_q}). 

\section{Conclusions}
This paper summarizes simulation results concerning the $\alpha$-relaxation of a dense, non-entangled polymer melt.  The melt was simulated by means of molecular-dynamics simulations, using a Lennard-Jones potential for modelling the monomer-monomer interactions and a FENE-potential for modelling the bonds.  The simulations were done at temperatures $0.46 \leq T \leq 0.7$ (measured in Lennard-Jones units), covering the temperature range from the onset of a two-step relaxation behaviour almost down to the critical temperature $\Tc \simeq 0.45$ of (ideal) mode-coupling theory (MCT).  We computed the incoherent, coherent chain, and coherent melt intermediate scattering functions to probe both single-monomer and collective dynamics of the melt. 

The $\alpha$-times $\tau^\mr{x}_q$ and the KWW-times $\tau^\mr{x K}_q$ are related to each other (see Eq.~(\ref{eq:tau_kww_and_alpha})) and exhibit similar features when $T$ approaches \Tc : a broad peak at wave vectors corresponding to the chains' radius of gyration and a sharp peak at the maximum of the collective static structure factor $S_q$ (Figures~\ref{fig:tau_q_and_S_q_vs_q_T046_T052_T07} and \ref{fig:tau_kww_inc+chain+melt_T0.47_vs_q}).  The peak at $q \approx 2 \pi / R_\mr{g}$ might perhaps be thought of as a consequence of the ``packing'' of the diffuse soft polymer coils in the melt, which should also contribute to the freezing of the system like that of the monomers does.  This peak is absent in simple liquids.  On the other hand, as in simple liquids, both time scales display a modulation that is in phase with $S_q$ for $q \gtrsim q_\mr{max}$. 

The quantitative analysis of the temperature dependence of the coherent $\alpha$-relaxation times shows that they follow a power law for $T \geq 0.47$ (Figure~\ref{fig:log_tau_alpha_chain+melt_vs_log_T-T_c}).  At $T=0.46$, however, there are clear deviations.  This implies that the temperature range, where idealized MCT is applicable, is left.  Moreover, at large wave numbers the power-law behaviour is progressively violated at $T \geq 0.65$ when $q$ increases.  This temperature marks an upper bound of the applicability of the aymptotic predictions of idealized MCT. The results show that the bound is $q$-dependent.  The exponent of the power law is only compatible with the MCT prediction from the $\beta$-analysis for wave vectors close to $q_\mr{max}$, but in general grows with $q$ for both incoherent and coherent scattering functions (Figure~\ref{fig:gamma_inc_chain_melt}).  This observation is (currently) not accounted for by MCT.  

On the other hand, the time-temperature superposition principle, an important prediction of MCT, is well borne out by the simulation data if $T \geq 0.47$.  For $T=0.46$, however, deviations from this scaling behaviour are encountered (Figure~\ref{fig:coh_melt_sf_rescaled_q3.0_q6.9_q14.0_allT}).  Within the framework of MCT such deviations are expected close to $\Tc$ due to relaxation processes which are not taken into account in the idealized version of the theory.  These processes change the shape of the correlators in the late-$\beta$/early-$\alpha$ regime \cite{FuchsGoetzeHildebrand1992_extMCT} so that a superposition with curves measured at higher temperatures is no longer possible.  However, MCT predicts that the additional relaxation processes do not perturb the scaling in the $\beta$-regime.  The factorization theorem should still hold \cite{FuchsGoetzeHildebrand1992_extMCT}.  Our simulation results provide evidence for this prediction (see Figure~5 of \cite{betaDynamics}). 

At the beginning of the $\beta$-regime oscillations were observed for the collective scattering functions at all $T$ which grow with decreasing $q$ (Figures~\ref{fig:coh_melt_sf_rescaled_q3.0_q6.9_q14.0_allT} and \ref{fig:coh_melt_sf_low_q_T0.47}).  Sound waves might be the origin of these oscillations, but a conclusive explanation is still to be found.  It is important to note, however, that these oscillation do (presumably) not alter the findings for the $\alpha$-relaxation presented here, since they already have no noticable influence on the major part of the $\beta$-process (see part I \cite{betaDynamics}).  

In a detailed analysis of the shape of the $\alpha$-decay by the Kohlrausch-Williams-Watts function two procedures were compared:  Firstly, we used three free fit parameters, and secondly, a fit with the KWW-prefactor set to the non-ergodicity parameter, thus fixing the KWW-time and only allowing the stretching exponent to vary.  The $q$-dependence of the collective KWW-stretching exponents $\beta^\mr{K}_q$ shows a modulation with $S_q$ independent of the fit method (Figure~\ref{fig:beta_K_coh_melt_free+fixed_NEP+S_q}).  Furthermore, the asymptotic prediction of MCT for the $q$-dependence of the KWW-times, $\tau^\mr{x K} \propto q^{-1/b}$, is verified for all correlators for large $q$ (Figure~\ref{fig:tau_kww_inc+chain+melt_T0.47_vs_q}). However, our data do not allow to clearly confirm the corresponding prediction for the stretching exponent in the same $q$-range.  In case of the coherent chain scattering function, a $q^{-3}$ dependence of the KWW-times at small $q$ is obtained. This behavior is compatible with recent experiments \cite{RichterMonkenbuschAllgeier1999}. 

In summary, one can say that ideal MCT provides a good starting point for the description of the $\alpha$-relaxation of our model if temperature is close, but not too close to $\Tc$.  We could give estimates for the temperature range of validity of the theory and for the $q$-range where corrections to MCT predictions become important.  There are, however, some polymer-specific effects, such as the strong increase of the $\alpha$-relaxation time on the length scale of the radius of gyration, which cannot be contained in a theory developed for simple liquids.  

To what extent the results discussed in this and the preceding paper are also observable in simulations of realistic polymer models is not clear at present.  An important step in this direction was made by the development of a model potential for polybutadiene \cite{GrantPaul1999,GrantPaul1998}, which yields very good agreement with neutron scattering experiments \cite{GrantPaul2000,GrantPaul_ChemPhys2000}.  A comparison of the dynamics of this model in the supercooled state with our data would represent an important test of our findings.  Results on a different model for polybutadiene \cite{ZonLeeuw1999} seem promising in this respect.  Therefore, we hope that our simulations can contribute to developing a detailed theory of the glassy behaviour of polymer melts.  Attempts in this direction are underway \cite{Guenza1999}.

\begin{acknowledgement}
We are indebted to C. Bennemann, J. Horbach, W. Kob, A. Latz, W. Paul, and F. Varnik for many helpful discussions and to M. Fuchs for valuable comments on the manuscript.  This work would not have been possible without generous grants of computing time by the HLRZ J{\"u}lich, the RHRK Kaiserslautern, the CTCMS at NIST, Gaithersburg, and the computer centre at the University of Mainz. Financial support by the Deutsche Forschungsgemeinschaft under grant No.\ SFB262 and by the ESF Programme on ``Experimental and Theoretical Investigation of Complex Polymer Structures'' (SUPERNET) is gratefully acknowledged.
\end{acknowledgement}

\bibliography{references_mct2}

\end{document}